\documentclass[twocolumn,prl,aps,superscriptaddress,showpacs]{revtex4}
\usepackage{xspace,amsmath,amsfonts,amsthm,amssymb,amsbsy,graphicx,color}

\begin{document}

\newtheorem{theo}{Theorem}
\newtheorem{lemma}{Lemma} 

\title{Energy Measurements and Preparation of Canonical Phase States of a Nano-Mechanical Resonator}

\author{Kurt Jacobs}

\affiliation{Department of Physics, University of Massachusetts at Boston, Boston, MA 02125, USA}

\affiliation{Hearne Institute for Theoretical Physics, Department of Physics and Astronomy, Louisiana State University, Baton Rouge, LA 70803, USA}

\author{Andrew N. Jordan}

\affiliation{Department of Physics and Astronomy, University of Rochester, Rochester, NY 14627, USA}  

\author{Elinor K. Irish} 

\affiliation{School of Mathematics and Physics, QueenÕs University Belfast, Belfast BT7 1NN, UK}

\begin{abstract}
We show that a continuous quantum non-demolition measurement of the energy of a nanomechanical resonator can be achieved by monitoring the resonator with a quantum point contact via a Cooper-pair box. This technique can further be used to prepare highly non-classical states of two resonators, such as canonical phase-reference states, and so-called ÒnoonÓ states. 
\end{abstract}

\pacs{85.85.+j,85.35.Gv,03.65.Ta,45.80.+r} 

\maketitle  

Nanoscopic mechanical resonators can now be built with frequencies in the hundreds of Megahertz and quality factors above $10^4$~\cite{Blencowe04, Huang03, Knobel03, LaHaye04, Naik06, Almog07}. These resonators have potential
applications in metrology~\cite{Ilic04} and information processing~\cite{Cleland04}, and hold the promise of realizing quantum behavior in bulk mechanical devices for the first time. To fully exploit the quantum properties of nanoresonators one must be able to prepare them in nonclassical states, but generating the necessary nonlinear evolution is challenging. Our purpose here is twofold. First we present a method for making a continuous QND (Quantum
Non-Demolition) measurement of the energy of a nano-resonator. Such measurements are important because they will prepare non-classical Fock states of the resonator, and allow the
observation of quantum jumps. While these measurements have been a problem of considerable interest for a number of years, only two methods to make such measurements in the solid-state exist to date~\cite{Santamore04,Buks06}, and only the latter, suggested recently, appears feasible with current technology. (We note that two methods to monitor energy in a near-QND fashion have also been proposed recently~\cite{Martin07,Jacobs07b}, as well as a promising QND scheme using an optical cavity~\cite{Thompson07}.) Second, we show that joint QND measurements of two resonators can be used
to prepare highly non-classical entangled states; in particular
canonical phase-reference (CP) states~\cite{Pegg88,Vaccaro03,Kitaev04} and so-called ``noon''  states~\cite{Kok02,Cable07}. Both can be thought
of as states of a single virtual oscillator whose number
basis consists of the states of definite phonon number-difference
between two oscillators. These states are very
difficult to create via Hamiltonian evolution, due to the
high-order nonlinearities required~\cite{Collett93}. Both CP states
and noon states have applications in high-precision phase
measurement. The former also contain a large store of entanglement,
and are therefore of potential use in information
processing. Further, when combined with a universal
procedure to create states of a single resonator, such as
that of Law and Eberly~\cite{Law96}, an energy measurement can
be used to create any state of the virtual oscillator. While
we will specifically consider nano-resonators in what follows,
the methods we describe here can just as readily
be applied to measurements and state-preparation of a
superconducting stripline resonator~\cite{Wallraff04}. Methods to entangle solid-state qubits using measurements have also been proposed (see {\it e.g.} Refs.~\cite{Mao04,Trauzettel06}).

To perform a QND measurement on a nano-resonator one requires an interaction Hamiltonian with a probe system proportional to $a^\dagger a$ where $a$ is the annihilation operator for the resonator. Such an interaction can be obtained quite easily by placing a Cooper-pair Box (CPB) adjacent to the resonator and ensuring that the detuning between them, $\Delta$, is large compared to their mutual interaction strength $\lambda$~\cite{Irish03}. The resulting interaction is $H_{\mbox{\scriptsize int}} = \hbar \mu \sigma_x a^\dagger a$, where the charge basis for the CPB is the eigenbasis of $\sigma_z$, and $\mu = \lambda^2/\Delta$. This kind of interaction has already been used to obtain spectral measurements revealing the discrete energy levels of a superconducting stripline resonator~\cite{Schuster07}.  

\begin{figure}[b] 
\leavevmode\includegraphics[width=0.75\hsize]{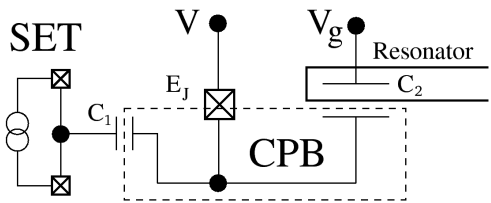}
\caption{The circuit diagram for the QND measurement scheme: $V$ and $V_g$ are applied voltages, and $C_1$ and $C_2$ are the capacitances connecting the three mesoscopic elements. The Cooper-pair box (indicated with a dashed box) is coupled to the nano-resonator and the single-electron transistor. } 
\label{fig0}
\end{figure}

To realize a continuous QND measurement we add one more element whose function is to perform a continuous measurement of the charge states of the CPB. This element can be either a quantum point-contact (QPC)~\cite{Korotkov} or single-electron transistor (SET) placed adjacent to the CPB. (A QPC is more efficient than an SET, but as we discuss below efficiency is not essential for an effective QND measurement). Monitoring the current through either of these devices provides a continuous measurement of $\sigma_z$.  The circuit diagram for the measurement scheme is shown in Fig.~\ref{fig0}. The full stochastic master equation (SME) for the coupled resonator-CPB system, under the continuous measurement is~\cite{Korotkov,JacobsSteck06} 
\begin{eqnarray}
  d\rho & = &  -i [H/\hbar,\rho]dt  - k[\sigma_z,[\sigma_z, \rho ]] dt  \nonumber \\ 
           &    &  + \sqrt{2k}[\sigma_z\rho + \rho\sigma_z  - 2\langle\sigma_z\rangle \rho] (dr - \langle\sigma_z\rangle dt)  ,
           \label{sme} 
\end{eqnarray} 
where  $H  = \hbar[ \omega_{\mbox{\scriptsize R}} a^\dagger a + \omega_{\mbox{\scriptsize C}} \sigma_z  +  \omega_{\mbox{\scriptsize J}}\sigma_x ] + H_{\mbox{\scriptsize int}}$ is the Hamiltonian,  
$\omega_{\mbox{\scriptsize R}}$ is the resonator frequency, $E_{\mbox{\scriptsize C}}$ is the CPB charging energy, and $\omega_{\mbox{\scriptsize J}}$ is it Josephson tunneling frequency. The parameter $k$ is the strength~\cite{DJJ} of the measurement of $\sigma_z$ and is determined by the coupling strength between the SET/QPC and the CPB. This master equation is stochastic because it is driven by the continuous random stream of measurement results $r(t)$. The increment of $r$ in time $dt$ is given by $dr = \langle\sigma_z\rangle dt + dW$~\cite{JacobsSteck06}, where $dW$ is a Gaussian noise increment satisfying the Ito calculus relation~\cite{WienerIntroPaper,JacobsSteck06}. 

\begin{figure}[t] 
\leavevmode\includegraphics[width=1\hsize]{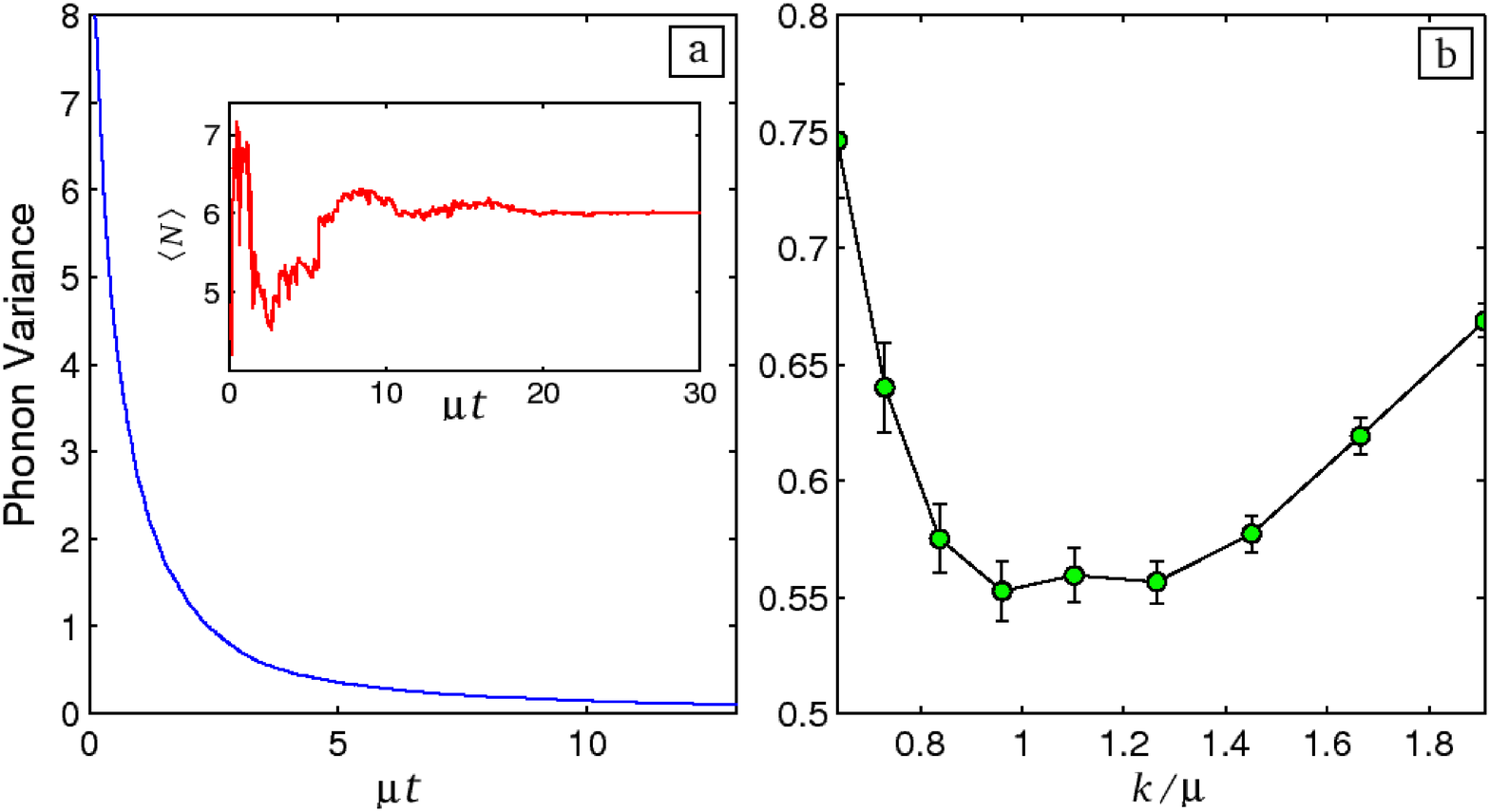}
\caption{(a) The evolution of the phonon-number variance of the resonator for $k=\mu$ averaged over a thousand measurements. Inset: The average value of the phonon number for a single measurement. (c) This average variance as a function of $k$ after measuring for a time $t=2\pi/\mu$.} 
\label{fig1}
\end{figure}

We show that this configuration realizes a continuous QND measurement of the energy of the resonator by simulating the master equation Eq.(\ref{sme}). Starting the resonator in a uniform superposition of the first ten number states, we find that the variance of the resonator's energy decays to zero and remains there as required, with the resonator projected onto a single number state. In Fig.~\ref{fig1}(a) we show the decay of the energy variance averaged over a thousand realizations.  It is worth noting that
the projection of the system onto an energy eigenstate is
unaffected by any environmental disturbance to the CPB
by virtue of the QND interaction. Decoherence can only
affect the measurement by reducing the information
extraction rate (by interfering with the rotation induced
by the resonator). However, one expects this to be significant
only if the decoherence rate is much greater than $\mu$. 

The operating principle behind the continuous measurement is that the effective rotation rate $\omega_x$ of the 
CPB about its $x$-axis depends directly on the number state of the resonator ($\omega_x= \omega_J + \hbar \mu a^\dagger a$). Because the SET-measurement is sensitive to the charge of the CPB ($\sigma_z$), it is also sensitive to the rate at which the $z$-eigenstate is changing, and hence the phonon number of the resonator. The measurement is QND by virtue of the CPB-resonator interaction.  The projected phonon number can be determined either from the full SET measurement record, or simply from the peak in its noise-spectrum~\cite{Korotkov01}.

The rate at which information is extracted from the resonator, being the rate at which the resonator is projected onto a phonon number state, is determined by the interaction strength with the CPB, $\mu$, and the measurement strength $k$. For a given value of $\mu$ we now wish to determine the value of $k$ that maximizes the projection rate. If we make $k$ too small then we expect this to limit the rate of information extraction. On the other hand, making $k$ too large will do the same: the information regarding the energy is contained in the rate at which the CPB flips between $z$-states, and the quantum Zeno effect induced by the measurement will prevent this flipping. There will therefore be an optimal value of $k$ for a given $\mu$. We find this value by calculating the energy variance after a fixed measurement time, and averaging this over many simulations of the measurement. The results are presented in Fig.~\ref{fig1}(b), showing that the optimal choice is $k\approx\mu$. Realistic parameters are $\mu = 10^6~\mbox{s}^{-1}$~\cite{Jacobs07b}, and $k$ can be much larger~\cite{Vandersypen04}. From Fig.~\ref{fig1}(a) we see that this gives a measurement time of about $20 \mu\mbox{s}$. 

\begin{figure}[t] 
\leavevmode\includegraphics[width=1\hsize]{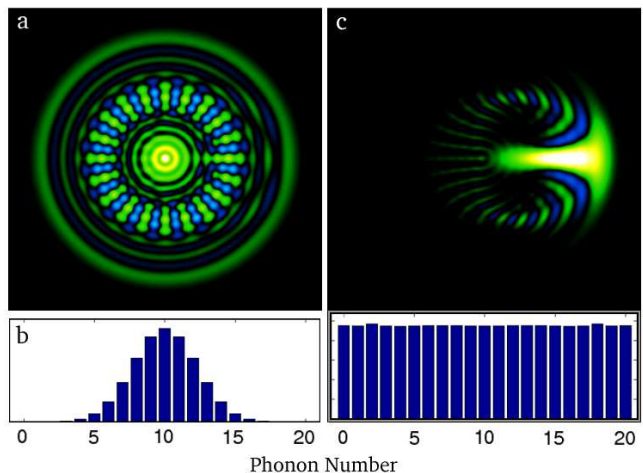} 
\caption{Wigner functions for (a) a ``noon'' state with $N=20$ and (c) a phase-state with  $N=20$, generated using a QND measurement. (c) Inset: phonon number-difference distribution for the same state. For these plots luminosity is proportional to the real part of the Wigner function. (b) Typical phonon-difference distribution generated when the initial states are coherent. } 
\label{fig2}
\end{figure}

We now show that energy measurements are a powerful tool for preparing highly non-classical entangled states. In particular we show that they can be used to convert single-oscillator Schr\"{o}dinger cat states, which are relatively easy to prepare, into two kinds of phase-reference states. This can be achieved with any detector capable of a joint-oscillator energy measurement.  For our measurement scheme, we place two nano-resonators symmetrically on either side of the CPB. In this case the CPB interacts with both resonators simultaneously, and the interaction Hamiltonian is $H_{\mbox{\scriptsize int}} = \mu\sigma_x(a^\dagger a + b^\dagger b)$, where $b$ is the annihilation operator for the second oscillator. The measurement now extracts information about the sum of the energies of the two oscillators. The measurement will therefore project the two oscillators into a subspace with a fixed total phonon number. If the states of the two resonators prior to the measurement are $|\psi_1\rangle = \sum_n c_n |n\rangle$ and $|\psi_2\rangle = \sum_n d_n |n\rangle$, respectively, then if the measurement projects onto the space with a total of $N$ photons, the combined state of the resonators becomes 
\begin{equation}
   |\psi\rangle \propto \sum_{n=0}^N c_n d_{N-n}  |n\rangle |N-n\rangle \equiv  \sum_{n=0}^N q_n  |n\rangle_{N}.
   \label{Estate}
\end{equation}
Each state in the sum, $|n\rangle_{N}$, is a state of definite phonon-number difference between the resonators.  If each resonator is prepared in the same (arbitrary!) coherent state prior to the measurement, then $|\psi\rangle$ is an entangled state with the ``N-binomial'' coefficients  $q_{n} \propto \sqrt{1/((N-n)!n!)}$. Fig~\ref{fig2}(b) shows a typical distribution $|q_n|^2$ for this case.  

We consider first preparing ``noon'' states of two oscillators, defined as $|\mathcal{N}\rangle = |N\rangle |0\rangle + |0\rangle |N\rangle$. These states are useful for very high precision measurements of a phase shift of one of the oscillators~\cite{Resch07} (the precision scales linearly with $N$). Consequently, there has been considerable recent work on methods to engineer these states in optical modes~\cite{Kok02,Cable07,Resch07}. 

One can prepare a noon state with arbitrary $N$ in the following way. We first prepare each oscillator in a superposition of two coherent states. Such a superposition is usually referred to as a Schr\"{o}dinger cat state. One of the coherent states is the vacuum, and the other, which we denote $|\alpha\rangle$, we choose so that its number-state distribution is peaked near $N$ phonons. Cat states are generated by applying the non-linear Hamiltonian $(a^\dagger a)^2$, and this could be done for a nano-resonator using the techniques in~\cite{Jacobs07x}. Note that so long as $N\gg 1$, the probability that each oscillator is the zero-phonon state is {\em much greater} than the probability that it is in the one-photon state. We now make a measurement of the combined phonon number of the two oscillators. If the result of the joint energy measurement is $N$ phonons, then, up to a normalization constant, the resonators are projected into the combined state 
\begin{equation*}
   |\psi\rangle \propto c_0 c_N  |\mathcal{N}\rangle + c_1c_{N-1}|1\rangle_N  +  c_2c_{N-2}|2\rangle_N\cdots ,
\end{equation*}
where we have used the fact that $d_n=c_n$. Since $c_{N-m} < c_N$ for $1 \leq m < N$, and in particular $c_m \ll c_0$ for $m \leq N/2$, the state is a noon state with high fidelity. This fidelity increases with $N$. If we wish to fix the value of $N$, we can decrease $c_1$, and thus increase the fidelity, by squeezing the non-vacuum coherent component. If we choose $\alpha  = \sqrt{N}$,  then the probability with which we obtain the desired measurement result is $P(N) \approx (2\pi N)^{-1/2}$. While this does decrease with $N$, it is quite reasonable for large values of $N$: for $N = 50$ the measurement succeeds 1 time in 17 tries. In addition, if one merely wants to produce a noon state with $N \geq N_{\mbox{\scriptsize min}}$, then one can achieve a near-unity success rate by choosing $\alpha \gg \sqrt{N_{\mbox{\scriptsize min}}}$. For the purposes of graphically representing the noon states we define a subsystem of the two oscillators whose basis states are indexed by the phonon-number difference between the two oscillators. This basis is the $|m\rangle_{N}$, with $m = 0,\ldots,N$. We can then plot the Wigner function for the state of this subsystem when the oscillators are in the noon state.  In Fig.~\ref{fig2}(a) we plot this Wigner function for the noon state generated using the above procedure when $\alpha = 6$ and the measurement projects onto the subspace with 20 phonons.  

We now turn to the task of preparing phase states. The problem of defining a phase operator conjugate to the number operator goes back to Dirac~\cite{Dirac27}, and was finally resolved in 1988 by Pegg and Barnett~\cite{Pegg88}. The phase eigenstates of the Pegg-Barnett phase operator (the canonical phase states) are given by $ |\theta\rangle \propto \lim_{N\rightarrow\infty} \sum_{0}^N e^{in\theta}|n\rangle$ for $\theta\in [0,2\pi]$. For a finite maximum phonon number $N$, there are $N+1$ orthogonal phase states, $|\theta_n\rangle$, corresponding to the phase angles $\theta_n = 2\pi n/N$. That is, when one is limited to a maximum excitation energy of $\hbar\omega N$, one can obtain a phase resolution of $\Delta\theta = 2\pi/N$. The zero-phase state for $N$-phonons is thus a real uniform superposition of all the number states from $n=0$ to $N$, and all the other $N$-phonons phase states are obtained from this by free evolution. 

Phase is not well-defined without a phase reference. While this statement is obvious for phase, it is, in fact, true for all physical observables, as was pointed out in the quantum context initially by Aharonov and Susskind~\cite{Aharonov67}. For any observable to be physically meaningful, one must have a second system, referred to as a {\em reference system}, or {\em reference frame}, with which one can compare the system containing the observable. In the last few years the subject of reference frames for quantum systems has been studied extensively~\cite{Kitaev04,Bartlett07}. If one has two oscillators, then the state-space of these two systems ``supports'' as a subsystem an oscillator with a well-defined phase. The number states of this oscillator are precisely the number-difference states $|m\rangle_{N}$ defined above, and the zero-phase phase state of this oscillator is thus $|\theta_0\rangle_{N} = \sum_{m=0}^N |m\rangle_{N} / \sqrt{N+1}$. Along with reference systems come the notion of a perfect {\em reference state}. This is a joint state of two systems, where the first is the base reference system,  such that the second system can be used as a perfect (surrogate) reference for measuring a third system~\cite{Vaccaro03}. The canonical phase state defined above, $|\theta_0\rangle_{N}$, is a perfect phase reference. It is also worth noting that of all the states spanned by the $|m\rangle_{N}$, it is the canonical phase states that realize the highest entanglement between the two resonators. Thus the problem of generating phase states is also that of maximizing the entanglement generated by a joint energy measurement. 

The difficulty of preparing phase states is due to the fact that their perfectly flat number distribution is truncated sharply at a fixed number $N$. However, 
the way in which a joint energy measurement folds together the states of the individual oscillators provides the ability to generate a flat distribution, and the subspace projection allows one to generate the sharp cut-off.  To begin we first recall how the joint energy measurement combines the coefficients $c_n$ and $d_n$ of the states of the two resonators (Eq.(\ref{Estate})). This tells us that we need to choose $c_n d_{N-n} = \mbox{const.}$, or if both oscillators have the same initial state, $c_n c_{N-n} = \mbox{const.}$ for some $N$. This is satisfied by setting $c_n \propto e^{-\beta n}$ for some constant $\beta$. We cannot arrange for this to be true exactly, but we can make it true to a good approximation by choosing the initial states of the oscillators to be a superposition of a small number of squeezed coherent states. As examples, it turns out that high accuracies can be achieved for $N=10$ and $N=20$ phase states with superpositions of only two and three squeezed states, respectively. Superpositions of multiple squeezed states can once again be obtained by applying the Hamiltonian $(a^\dagger a)^2$ for specified times. (We choose all the states to have the same squeezing, as this simplifies preparation~\cite{Jacobs07x}). The more coherent states in the initial superposition, then the higher the fidelity with which we can generate a phase states for a given $N$. 

Defining the error in the preparation as $f = 1 - |\langle \psi| \theta_0 \rangle_N|^2$, where $|\psi\rangle$ is the state we prepare, we find that using a superposition of two mildly squeezed states we can prepare an $N=10$ phase state with an error of less that $10^{-5}$, and with three squeezed states we can prepare an $N=20$ phase state with $f = 1.2\times10^{-5}$. To obtain these results we performed a numerical optimization over initial states. For generating an $N=10$ phase state the optimal amplitudes of the squeezed components are $\alpha_1 = 1.162$, $\alpha_2 = 3.277$ with a (very mild) squeezing parameter of $s=-0.097$. For an $N=20$ phase state these are $\alpha_1 = 1.241$, $\alpha_2 = 3.100$, $\alpha_3 = 5.024$ with a mild squeezing parameter $s=-0.1131$. For higher values of $N$ one requires more coherent components in the initial state in order to maintain the same fidelity. 

As a final note, it is clear from Eq.(\ref{Estate}) that joint-energy QND measurements are universal for preparing states of the virtual oscillator (entangled states whose Schmidt basis is $\{|n\rangle_N\}$) when combined with a universal technique for preparing states of a single oscillator. That of Law and Eberly~\cite{Law96} would be suitable for this purpose when $N$ is not too large.  

\vspace{-0.3em}

{\em Acknowledgements:} K.J. was supported by the Army Research Office and the Disruptive Technologies Office. E.I. was supported by the UK EPSCR. 


\end{document}